\documentclass[12pt]{article}
\usepackage{amssymb}
\usepackage{euscript}
\usepackage{epsfig,bbm}
\def\EN{\EuScript{N}}
\def\EF{\EuScript{F}}
\def\EV{\EuScript{V}}

\def\d{\partial}

\def\ln{{\rm ln}}
\def\a{\alpha}
\def\0{\nonumber}
\def\ee{\end{eqnarray}}     %eqnarray
\def\be{\begin{eqnarray}}
\def\ba{\begin{array}}          %array
\def\ea{\end{array}}

\begin{document}
\begin{flushright}
{SISSA 66/2005/EP}\\
{hep-th/0511006}
\end{flushright}

\begin{center}
{\LARGE {\bf Light--cone Superstring Field Theory,
}}\\[0.3cm]
{\LARGE {\bf
pp--wave background and
}}\\[0.3cm]
{\LARGE {\bf integrability properties}}

\vskip 1cm

{\large L. Bonora$^a$, R.J.Scherer Santos$^b$, A.S. Sorin$^c$ and D.D.Tolla$^b$}
{}~\\
\quad \\
{\em ~$~^{a}$International School for Advanced Studies (SISSA/ISAS),}\\
{\em Via Beirut 2, 34014 Trieste, Italy and INFN, Sezione di Trieste}\\
 {\tt bonora@sissa.it}
{}~\\
\quad \\
{\em ~$~^{b}$International School for Advanced Studies (SISSA/ISAS),}\\
{\tt scherer@sissa.it \quad tolla@sissa.it}
{}~\\
\quad \\
{\em ~$~^{c}$Bogoliubov Laboratory of Theoretical Physics,}\\
{\em Joint Institute for Nuclear Research (JINR),}\\
{\em 141980, Dubna, Moscow Region, Russia}\\
{\tt sorin@theor.jinr.ru}

\end{center}

\vskip 2cm {\bf Abstract.} {We show that the three strings vertex coefficients
in light--cone open string field theory satisfy the Hirota equations for the
dispersionless Toda lattice hierarchy. We show that Hirota equations
allow us to calculate the correlators of an associated quantum system where the Neumann
coefficients represent the two--point functions. We consider
next the three strings vertex coefficients of the light--cone string field
theory on a maximally supersymmetric pp--wave background. Using the previous
results we are able to show that these Neumann coefficients satisfy
the Hirota equations for the full Toda lattice hierarchy at least up to
second order in the 'string mass' $\mu$.}

\vskip 1cm

\section{Introduction}

In recent years it has become more and more evident that integrability plays
an important role in string theory. This fact is conspicuous in
the ${\rm AdS_5\times S^5}$ example of AdS/CFT duality, where integrability
features on both sides of the correspondence. Integrability plays a major role
in connection with topological strings. But there are also other less
well--known cases. One of these is represented by the integrability properties
of the Neumann coefficients for the three
strings vertex of string field theory (SFT).
In \cite{BS} it was shown that these Neumann coefficients in the case of
Witten's covariant open bosonic string field theory (OSFT), \cite{W1},
satisfy the Hirota equations, \cite{TL}, of the {\it dispersionless}
Toda lattice hierarchy \cite{TT,CD,Zabrodin}.
This is to be traced back to the existence of a conformal mapping \cite{LPPI,BR}
underlying the three strings vertex. The conclusions of \cite{BS} were limited
to covariant OSFT. What we would like to show in this paper is that
the above integrability properties seem to be a general characteristic of
the three strings vertex. Indeed, below, we will
prove that it holds for the light--cone string field theory (LCSFT) in
flat background. But, what is more important, we will present evidence that
it may also hold for a nontrivial background. We will specifically
examine the Neumann coefficients of the three strings
vertex in a maximally supersymmetric pp--wave background and expand them
in terms of the `string mass' $\mu$. We will show that, up to second order in
this expansion, they satisfy the (correspondingly expanded) Hirota equations
for the {\it dispersive} Toda lattice hierarchy (i.e. for the full Toda lattice
hierarchy), \cite{UT}. This leads us to the conjecture that the Neumann
coefficients
in this background satisfy the Hirota equations of the full Toda lattice
hierarchy.

The results of this paper are admittedly of mere theoretical interest. They do
not have an immediate practical impact. And perhaps they are not unexpected.
Once the generating function of the Neumann coefficients are written in terms
of conformal
mappings (see below), they are expected to satisfy some kind of integrability
condition, \cite{BR}. That this is true also for the three strings vertex in the
pp--wave background may be seen as the consequence of the solvability
of string theory on such background. However the main value of our results is
that they point towards the existence of an integrable model (of which the Hirota
equations are a signal) underlying the
SFT structure, very likely a matrix model, which, once uncovered, would greatly
improve our knowledge of SFT. The fact that this is probably true also for
a nontrivial background, such as the pp--wave one, may suggest a way to approach
the problem of defining SFT on more general backgrounds.

This paper is organized as follows. In section 2 we introduce the Neumann
coefficients for LCSFT, define their generating functions and the conformal
maps they can be derived from, and
study their properties. In section 3 we show that these generating
functions satisfy the Hirota equations for the dispersionless Toda lattice
hierarchy. In section 4 we define the Hirota equations for correlators with
more than two insertions. In section 5 we introduce the Neumann coefficients
in a pp-wave background and prove that they satisfy the Hirota equations for the
full Toda lattice hierarchy up to second order in $\mu$.

\section{The conformal maps for the light--cone SFT}

In this section we work out the conformal properties of the three strings vertex
in light--cone SFT. We introduce first the relevant Neumann coefficients and
then we show that, like in Witten's OSFT, they can be defined in terms of
conformal mappings and determine them explicitly. We show how our formulas
are related to the ones of Mandelstam, \cite{Mandel}.  Finally we discuss the
general features of these conformal mappings and describe an explicit
example.

\subsection{Neumann coefficients for bosonic three strings vertex}

In LCSFT the three superstrings vertex is determined entirely in terms of the
Neumann coefficients of the bosonic part. Therefore we will concentrate
on the latter, which can be written in the form
\be
|V_3\rangle = \int\prod_{s=1}^3 \,
d\a_s \prod_{I=1}^8 dp_{(r)}^I \delta(\sum_{r=1}^3\a_r)
\delta(\sum_{r=1}^3 p_{(r)}^I) {\rm exp}(-\Delta_B)\, |0,p\rangle_{123}
\label{V3}
\ee
where $I=1,\ldots,8$ label the transverse directions, $\a_r =2p_{(r)}^+$ and
\be
\Delta_B = \sum_{r,s=1}^3 \delta_{IJ}\left(\sum_{m,n\geq 1}a_m^{(r)I\dagger}
V_{mn}^{rs}a_n^{(s)J\dagger} + \sum_{n\geq 1} p_{(r)}^I
V_{0n}^{rs}a_n^{(s)J\dagger}
+ p_{(r)}^I V_{00}^{rs}p_{(s)}^J\right) \label{DeltaB}
\ee
where summation over $I$ and $J$ is understood.
The operators $a_n^{(s)J},a_n^{(s)J\dagger}$ are the non--zero mode
transverse oscillators of the $s$--th string. They satisfy
\be
[a_m^{(r)I},a_n^{(s)J\dagger}]=\delta_{mn}\delta^{IJ}\delta^{rs}\0
\ee
$p_{(r)}$ is the transverse momentum of the $r$--th string and
$|0,p\rangle_{123}\equiv |p_{(1)}\rangle\otimes
|p_{(2)}\rangle\otimes |p_{(3)}\rangle$ is the tensor product of the Fock vacua
relative to the three strings. $|p_{(r)}\rangle$ is annihilated by $a_n^{(r)}$
and is the eigenstate of the operator $\hat p_{(r)}^I$ with eigenvalue
$p_{(r)}^I$. The vertex coefficients $V_{nm}$ are more conveniently expressed
in terms of the Neumann coefficients $N_{nm}$ as follows:
\be
&&V_{mn}^{rs}=- \sqrt{mn} N_{mn}^{rs},\quad\quad\quad V_{m0}^{rs}=
\frac{\sqrt{m}}{6} \epsilon^{stu} \frac{\a_t-\a_u}{\a_r} N_m^r\equiv
\frac 1{\sqrt{m}} N_{m0}^{rs}\label{VN}\\
&&V_{00}^{rr}=\frac 13 (\a_{r-1}^2 + \a_{r+1}^2)\frac {\tau_0}\a,\quad\quad\quad
V_{00}^{rs}= -\frac 23 \a_r\a_s\frac {\tau_0}\a,\quad\quad r\neq s~. \label{V00}
\ee
In these equations and in the sequel $r,s=1,2,3$ modulo 3. In
eq.(\ref{VN}) $\epsilon$ denotes the completely antisymmetric tensor with
$\epsilon^{123}=1$ and the indices $t,u$ are summed over.
 In these equations the $N$'s are related to the Mandelstam $\bar N$
 by $N_{mn}^{rs}=\bar N_{mn}^{rs}$ and
 $N_m^r=\a_r \bar N_m^r$, with, see
\cite{Mandel,GS},
\be
&& N^{r}_m =  \frac{\Gamma(-\frac{\alpha_{r+1}}{\alpha_r}~ m)}{m!~
\Gamma(-\frac{\alpha_{r+1}}{\alpha_r}~m-m+1)}~e^{\frac{\tau_0}{\alpha_r}m},
\label{Nr-1}\\
&&N^{rs}_{mn} =-\frac{\alpha~ m~n}
{\alpha_r \alpha_s (m~\alpha_s+n~\alpha_r)}
N^{r}_{m}N^{s}_{n}
\label{Nrs-2}
\ee
and
\be
\a = \a_1\a_2\a_3,\quad\quad \tau_0=
\sum_{r=1}^{3}\alpha_r ~\ln~|\alpha_r|,\quad \quad
\sum_{r=1}^{3}\alpha_r=0~.
\label{const1}
\ee
The three strings vertex $V_{m0}^{rs}$ and $V_{00}^{rs}$ are not exactly the
same as in \cite{GS}. We have symmetrized them using momentum
conservation.

The Neumann coefficients (\ref{Nr-1}--\ref{Nrs-2}) are homogeneous
functions of the parameters ${\alpha_r}$,
so they actually depend on a single parameter $\beta$
\be
&& \beta_r:= \frac{\alpha_{r+1}}{\alpha_r}, \quad \beta_1\beta_2\beta_3=1,
\quad \beta_r (\beta_{r+1}+1)=-1,
\label{ident}\\
&&\beta_1=\beta,
\quad \beta_2=-\frac{\beta+1}{\beta}, \quad
\beta_3=-\frac{1}{\beta+1},\label{betar}\\
&& -\infty < \beta_r \leq -1,\quad  -1 \leq \beta_{r+1} \leq 0, \quad
0 \leq \beta_{r+2} < \infty, \0\\
&&e^{\frac{\tau_0}{\alpha_1}}=
\frac{|\beta|^\beta}{|\beta+1|^{\beta+1}},\quad
e^{\frac{\tau_0}{\alpha_2}}=
\frac{|\beta|}{|\beta+1|^{\frac{\beta+1}{\beta}}},\quad
e^{\frac{\tau_0}{\alpha_3}}=
\frac{|\beta+1|}{|\beta|^{\frac{\beta}{\beta+1}}}~.
\label{const2}
\ee

We can organize the Neumann coefficients by means of generating functions
\be
&&N^{r}(z) := \sum_{n=1}^\infty \frac 1{z^n} N^{r}_{n}, \quad\quad
N^{rs}(z_1,z_2) := \sum_{n,m=1}^\infty
\frac 1{z_1^n} \frac 1{z_2^m} N^{rs}_{nm}~.
\label{Nrs-gen}
\ee
Our first purpose is to express these functions in terms of conformal mapping
from the unit semidisk to the complex plane, in analogy the case of covariant
OSFT, \cite{BS}.

We write the conformal mappings for LCSFT as follows
\be
f_r(z^{-1}) = f_{r+2}(0)
-\frac{(f_{r+2}(0)-f_r(0))(f_{r+1}(0)-f_{r+2}(0))}
{(f_{r+1}(0)-f_r(0))\varphi_r(z^{-1})+f_{r}(0)-f_{r+2}(0)}\label{fi}
\ee
with $f_1(0)\neq f_{2}(0)\neq f_{3}(0)$. The functions
$\varphi_r(z^{-1})\equiv \varphi_r$ are solutions
to the equations
\be
\varphi^{\beta_r}_r(\varphi_r-1)
=\frac{1}{z}e^{\frac{\tau_0}{\alpha_r}}:= x_r~.
\label{eqs-y}
\ee
We remark that, as a consequence of (\ref{fi}--\ref{eqs-y}),
we have the following identities
\be
&&f_r'(0)=
e^{\frac{\tau_0}{\alpha_r}}\frac{(f_{r+1}(0)-f_r(0))(f_{r+2}(0)-f_r(0))}
{f_{r+1}(0)-f_{r+2}(0)}, \nonumber\\
&&\frac{f_r'(0) f_{s}'(0)}{(f_r(0) - f_{s}(0))^2} =
e^{\tau_0(\frac{1}{\alpha_r}+ \frac{1}{\alpha_{s}})}, \quad r\neq s~.
\label{f}
\ee
By means of the conformal mappings we now define the generating functions
\be
N^{r}(z)\!&=&\!-\beta_{r+2}~ \ln
\left( -\frac{f_r'(0)}{z} \, \left(\frac 1{f_r(0) -
f_r(z^{-1})}+ \frac 1
{f_{r+1}(0)-f_r(0)}\right)\right),\label{Nr}\\
N^{rr}(z_1, z_2)\!&=&\! \ln
\left( \frac{f_r'(0)}{z_1-z_2} \, \left(\frac 1{f_r(0) -
f_r(z_2^{-1})}- \frac 1
{f_r(0)-f_r(z_1^{-1})}\right)\right),\label{Nrr}\\
N^{rs}(z_1, z_2) &=& \ln \left( \frac {(f_r(z_1^{-1}) - f_s
(z_2^{-1})) (f_r(0)- f_s(0))} { (f_r(0)- f_s (z_2^{-1}))
(f_r(z_1^{-1})- f_s(0))} \right), \quad r\neq s~. \label{Nrs}
\ee
We notice that these definitions are very close to those in
Witten's SFT, \cite{BS}.

Let us analyze eq. (\ref{eqs-y}). First of all we notice that any solution will
have a branch that can be expanded around the value 1 as $1+x_r+\ldots$
\be
\varphi_r \!&\equiv&\! \varphi_r(z^{-1})\equiv \varphi_r(x_r)  =1 + x_r
+\sum_{k=2}^{\infty}a_k x_r^k\nonumber\\
\!&=&\! 1 +x_r -\beta_r x_r^2
+\frac{1}{2}\beta_{r}(1+3\beta_{r}) x_r^3
-\frac{1}{3}\beta_{r}(1+2\beta_{r})(1+4\beta_{r})x_r^4\nonumber\\
\!&+&\! \frac{1}{24}\beta_{r}(1+5\beta_{r})
(2+5\beta_{r})(3+5\beta_{r})
x_r^5 + ...\quad .\label{y_r}
\ee
Using these expansions inside the definitions (\ref{Nr}--\ref{Nrs})
one can make a direct comparison and verify that they do
generate the Neumann coefficients of Mandelstam (\ref{Nr-1}--\ref{Nrs-2}).

It must be clarified that the three functions $\varphi_1,\varphi_2,\varphi_3$
are not distinct.
With very simple manipulations one can see that the solutions
to eqs.(\ref{eqs-y}) with different $r$ are related by the following
transformations:
\be
\varphi_{r+1}\left(x_{r+1}\right)=
\frac{1}{1 - \varphi_r\left(-x_{r+1}^{\beta_r}\right)}
=1-\frac{1}
{\varphi_{r-1}\left((-x_{r+1})^{\frac{1}{\beta_{r+1}}}\right)}.
\label{trans-y}
\ee
Therefore the three equations (\ref{eqs-y}) give rise to a unique solution
(which describes a Riemann surface, see below).

The importance of these transformations should not be underestimated.
The three strings vertex represents the fusion of two strings that come together
and give rise to a third one. In Witten's covariant OSFT this process is very
symmetric: two strings evolving from $\tau=-\infty$ come together at $\tau=0$
in such a way that the right half of one string overlap with the left half of
the other; the three strings vertex describes this process which ends
with the emergence of the third string. The three strings vertex in the present
case (LCSFT), as we shall see, has a different stringy/geometric
interpretation. It is however still characterized by precise overlapping
conditions that are made possible by the above equations.

Let us analyze the gluing conditions for the mappings $f_r(z_{r}^{-1})$
 (\ref{fi}),
i.e. let us see how we can satisfy the conditions
\be
f_{r+1}(z_{r+1}^{-1}) = f_{r}(z_{r}^{-1})~.
\label{glue}
\ee
They lead to
\be
\varphi_{r+1}\left(x_{r+1}\right)=
\frac{1}{1 - \varphi_r\left(x_{r}\right)}
\label{cond}
 \ee
which can easily be solved if one compares eqs. (\ref{cond}) and
(\ref{trans-y})
\be
x_{r}=-x_{r+1}^{\beta_r}.\label{solut1}
 \ee
Equivalently, we have the following relations for the string coordinates:
\be
z_{r}=-z_{r+1}^{\beta_r}~.
\label{solut2}
 \ee
For later use we record that on the unit circle these conditions become
\be
z_{r}=e^{i \theta_r}, \quad \theta_{r}= \beta_r~ \theta_{r+1}+\pi.
\label{solut3}
 \ee
Starting from the definitions (\ref{Nr}--\ref{Nrs}) and using
(\ref{ident}), (\ref{fi}) and (\ref{eqs-y}),
one can derive the following explicit representations
\be
N^{r}(z)&=&   \frac{1}{\beta_{r}}~ \ln
\left( -\frac{f_r'(0)}{z} \, \left(\frac 1{f_r(0) -
f_r(z^{-1})}+ \frac 1
{f_{r+2}(0)-f_r(0)}\right)\right)\nonumber\\
&\equiv&  \ln ~\varphi_r(z^{-1})\equiv
-\beta_{r+2}~ \ln
\left(-\frac{e^{\frac{\tau_0}{\alpha_r}}}{z} \,
\left(\frac 1{1 - \varphi_r(z^{-1})}-1\right)\right)\nonumber\\
&\equiv& \frac{1}{\beta_{r}} ~\ln
\left(-\frac{e^{\frac{\tau_0}{\alpha_r}}}{z} \,
\frac 1{1 - \varphi_r(z^{-1})}\right),
\label{Nr1}
 \ee
 \be
N^{rr}(z_1, z_2)&=&  \ln \left( \frac{1}{z_1-z_2} \, \left(
z_1 \varphi_r^{\beta_r}(z_1^{-1}) -
z_2 \varphi_r^{\beta_r}(z_2^{-1})\right)\right)\nonumber\\
&\equiv& \ln
\left( \frac{e^{\frac{\tau_0}{\alpha_r}}}{z_1-z_2} \, \left(\frac 1{1 -
\varphi_r(z_2^{-1})}- \frac 1
{1-\varphi_r(z_1^{-1})}\right)\right),
\label{Nrr1}
 \ee
\be
N^{rr+1}(z_1, z_2) &=& \ln \left(\frac{1}{\varphi_r(z_1^{-1})}
+{\varphi_{r+1}(z_2^{-1})-
\frac{\varphi_{r+1}(z_2^{-1})}{\varphi_{r}(z_1^{-1})}} \right)
\nonumber\\ &\equiv&
\ln \left(1+\frac{(1-\varphi_r(z_1^{-1}))(1-
\varphi_{r+1}(z_2^{-1})}{\varphi_{r}(z_1^{-1})} \right),
\nonumber\\
&&N^{rs}(z_1,z_2)= N^{sr}(z_2,z_1)
\label{Nrs1}
\ee
which actually do not depend on the values of the conformal mappings
(\ref{fi}) at the origin, $f_{r}(0)$, as long as they are distinct, i.e.
$f_1(0)\neq f_{2}(0)\neq f_{3}(0)$. Therefore the parameters $f_{r}(0)$
in (\ref{fi}) are inessential gauge parameters.

\subsection{A comparison with the Neumann function method}

Eqs. (\ref{eqs-y}) are related to the ones that appear in the
derivation of open string tree amplitudes by means of the Neumann function
method. At the tree level the interaction of open strings consists
of the joining of two strings by the endpoints to form a unique string or
the splitting of a string into two. These two processes can
geometrically be described, \cite{Mandel,GSW,Kaku}, in terms of cutting and
pasting string
world--sheet strips. In turn such a geometry can be
nicely represented in the complex plane by means of logarithmic conformal
mappings. This leads to the explicit evaluation of the Neumann coefficients
for the interaction of three strings that are precisely those introduced at
the beginning of section 2. For instance, a suitable logarithmic map
for the $r$--th three--string configuration is
\be
\rho_r= \a_{r+1} {\rm ln}(z'-1) +\a_{r-1} {\rm ln} z'\label{rho}
\ee
where $z'$ takes values in the upper half plane. This can be transformed into,
see \cite{GSW,Kaku},
\be
y_r=- \beta_r {\rm ln} (1+x_r e^{y_r})~.
\label{ymand}
\ee
This equation can easily be reduced to the form (\ref{eqs-y}) if we
set $\varphi_r={\rm exp}(-\beta_{r+1}\beta_{r+2}y_r)$ and make the
identification
\be
x_r\equiv \frac {e^{\frac{\tau_0}{\a_r}}}z =
- z'^{\frac 1{\beta_{r-1}}}(z'-1)^{\beta_r}~.
\label{x3x'}
\ee

This means that our variable $z$ in (\ref{eqs-y}) is a `uniformizing' variable
for the three equations.
To clarify this issue, in the next subsection, we work out an explicit example.

\subsection{Particular case $\beta=1$}

 Let us study in detail a particular case, specified by values of $\a_r$
for which the parameter $\beta=1$. This case turns out to be particularly
simple and can be analyzed in full detail. In this case $\beta_1=1, \beta_2=-2,
\beta_3=-\frac 12$ and eqs. (\ref{eqs-y}) can easily be solved. For each
$\varphi_r$ we have two branches
\be
\varphi_1^\pm(z^{-1})&=&\frac{1\pm\sqrt{1+\frac{1}{z}}}{2},
\label{y-part0}\\
\varphi_2^\pm(z^{-1})&=& \frac{2}{1\pm\sqrt{1-\frac{1}{z}}},
\label{y-part1}\\
\varphi_3^\pm(z^{-1})&=&\left(\frac{1}{z}\pm\sqrt{1+\frac{1}{z^2}}\right)^2~.
\label{y-part}
\ee
The $^{(+)}$ branch is the one for which the expansion for small $x_r$ is of
the form $\varphi_r= 1+x_r +\ldots$.
We can now define the corresponding conformal mappings. To specify the latter
we need to fix the values they take at the origin. We recall that these
values can be arbitrary as long as they are distinct. Therefore we choose them
simply as
\be
f_1(0) = -1,\quad\quad f_2(0)=0,\quad\quad f_3(0)=1~.\0
\ee
With this choice we get
\be
&&f_1^{(\pm)}(1/z) = - \frac{1\pm\sqrt{1+1/z}}{3\mp\sqrt{1+1/z}}~,\label{f1}\\
&&f_2^{(\pm)}(1/z) = \frac{\sqrt{1-1/z}\mp 1}{\sqrt{1-1/z}\pm 3}~,\label{f2}\\
&&f_3^{(\pm)}(1/z) = -\frac 1{1-2(1/z \pm \sqrt{1+(1/z)^2})^2}~.
\label{f3}
\ee
Each couple of functions $f_r^{(\pm)}(1/z)$ defines a Riemann surface
represented
by two sheets joined through a cut. In the first case the cut runs from $-1$ to
$\infty$, in the second it runs from 1 to $\infty$ and in the third between
$i$ and $-i$. These three Riemann surfaces however are related by the maps
(\ref{trans-y}).

In the $\beta=1$ case it is rather easy to see in detail the gluing conditions
that characterize the three strings vertex. We consider three unit
semidisks $S_1,S_2$ and $S_3$ cut out in the complex
$\zeta_1=1/z_1$, $\zeta_2=1/z_2$ and
$\zeta_3=1/z_3$ upper half planes. Each function
$f_r^{(\pm)}$ maps the semidisk into a region $D_r^{(\pm)}$ in the image
complex plane. $D_1^{(\pm)}$ and $D_2^{(\pm)}$ have the form of lobes,
while $D_3^{(\pm)}$ are the outer part of a compact bi--lobed domain
in the lower (+) and upper (--) half plane, respectively (see Fig. 1--3,
where the $D_r^{(+)}$'s are shown, the $D_r^{(-)}$'s being the same
regions reflected with respect to the real axis).
\begin{figure}[htbp]
    \hspace{-0.5cm}
\begin{center}
    \includegraphics[scale=1]{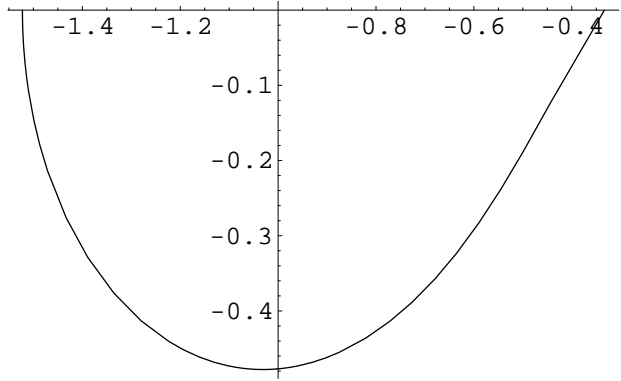}
    \end{center}
\caption{\emph{\small The region $D_1^{(+)}$  is contained between the curve
and the real axis.}}
    \label{fig:A}
\end{figure}

\begin{figure}[htbp]
    \hspace{-0.5cm}
\begin{center}
    \includegraphics[scale=1]{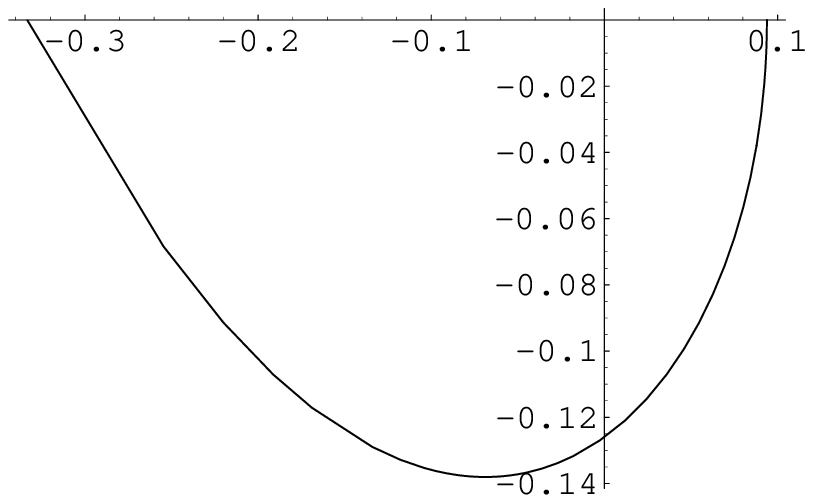}
    \end{center}
\caption{\emph{\small The region $D_2^{(+)}$ is contained between the curve and
the real axis.}}
    \label{fig:B}
\end{figure}

\begin{figure}[htbp]
    \hspace{-0.5cm}
\begin{center}
    \includegraphics[scale=1]{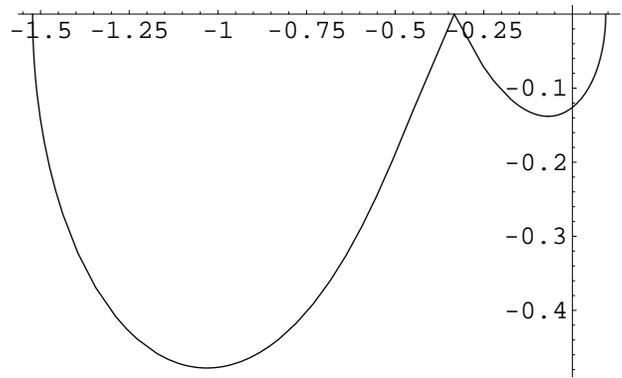}
    \end{center}
\caption{\emph{\small The region $D_3^{(+)}$ is the portion of the lower half
plane external to the curve.}}
    \label{fig:C}
\end{figure}

These domains are glued together along the borderlines in the
following way. To start with, $ f_1^{(+)}(\zeta_1)=f_2^{(-)}(\zeta_2)$
if $\zeta_1=-\zeta_2$.
This means that $S_1$ and $S_2$ are glued together along
the real axis with opposite orientation. The image common boundary
in $D_1^{(+)}$ and $D_2^{(-)}$ stretches also along the real axis.
The second overlap condition we consider is between $D_1^{(+)}$ and $D_3^{(+)}$:
$ f_1^{(+)}(\zeta_1)=f_3^{(+)}(\zeta_3)$ if $\zeta_3= - 1/\sqrt{\zeta_1}$. This
means the absolute values of $\zeta_1$ and $\zeta_3$ are 1, while their phases
$\theta_1$ and $\theta_3$ are related by $\theta_3 = \pi-\frac {\theta_1}2$.
So, while $\theta_1$ runs between 0 and $\pi$, $\theta_3$ spans the interval
between $\pi$ and $\pi/2$. These two angular intervals are mapped to the same
curve in the target complex plane: the curved part of the lobe boundary in
$D_1^{(+)}$ and a piece of the curved boundary of $D_3^{(+)}$ . Finally
$ f_2^{(+)}(\zeta_2)=f_3^{(+)}(\zeta_3)$ if $\zeta_2= - 1/\zeta_3^2$. That means
the moduli of $\zeta_2$ and $\zeta_3$ are 1, while their phases
$\theta_2$ and $\theta_3$ are related by $\theta_2 = \pi-2\theta_3$.
So, while $\theta_2$ runs between 0 and $\pi$, $\theta_3$ spans the interval
bewteen $\pi/2$ and $0$. Again these two angular intervals are mapped
to the same curve in the target complex plane: the curved boundary
of $D_2^{(+)}$ and the other piece of the curved boundary of $D_3^{(+)}$. In a similar
way one can proceed with the remaining three correspondences.

So far the discussion has been purely mathematical, without any concern for
the string process we want to describe. From the point of view of string theory
the `physical' branches are the ones where the expansion for small $x_r$ is of
the form $\varphi_r= 1+x_r +\ldots$, i.e. the + branches in
(\ref{y-part0}--\ref{y-part}).
So we have to glue together the images of the semidisks $S_1,S_2,S_3$ by
$f_1^{(+)},f_2^{(+)},f_3^{(+)}$ respectively, according to the maps
(\ref{solut2}), identifying the common boundaries. This means that we have to
glue the inner part of the lobe $D_1^{(+)}$ and  $D_2^{(+)}$ with $D_3^{(+)}$,
i.e. with the outer part of the bi--lobe. The result is the entire lower
half plane.
The process described is the joining of the strings 1 and 2 by two endpoints
to form the third string at the interaction point $\zeta_1=-1, \zeta_2=1$
and $\zeta_3=i$. The picture
obtained in this way allows us to make a comparison with the three string
interaction in covariant SFT. As one can see there are remarkable differences.
In particular the midpoints of the three strings never overlap (at variance
with covariant OSFT); on the contrary, the LCSFT vertex preserves its
characteristic perturbative geometry where the strings interact by the
endpoints.

For completeness we write down the explicit expressions of the
generating functions for the Neumann coefficients.
In these definitions we always use the + branch of $\varphi_r$
\be
N^{1}(z)&=& \ln \frac{1+\sqrt{1+\frac{1}{z}}}{2},\quad
N^{2}(z)=-N^1(-z), \quad
N^3(z)=2~\ln\left(\frac{1}{z}+\sqrt{1+\frac{1}{z^2}}\right)~~~~~
\label{Nr-part}
 \ee
 and
\be
&&N^{11}(z_1, z_2) = \ln \left( \frac{1}{2(z_1-z_2)}
\left(\frac 1{1-\sqrt{1+\frac{1}{z_2}}}
-\frac 1{1-\sqrt{1+\frac{1}{z_1}}}\right)\right),\0\\
&&N^{22}(z_1, z_2)=N^{11}(-z_1, -z_2), \0\\
&&N^{33}(z_1, z_2)=\ln \left( \frac{1}{z_1-z_2}
\left(\sqrt{1+z^2_1}-\sqrt{1+z^2_2}\right)\right)
\label{Nrr-part}
 \ee
as well as
\be
&&N^{12}(z_1, z_2)=-N^{11}(z_1, -z_2), \0\\
&&N^{23}(z_1, z_2) =
\ln \left(\frac{\sqrt{1-\frac{1}{z_1}}-\sqrt{1+z^2_2}}
{1-\sqrt{1+z^2_2}}\right),\0\\
&&N^{13}(z_1, z_2) =
\ln \left(\frac{\sqrt{1+\frac{1}{z_1}}+\sqrt{1+z^2_2}}
{1+\sqrt{1+z^2_2}}\right), \quad N^{rs}(z_1,z_2)= N^{sr}(z_2,z_1).
\label{Nrs-part}
 \ee

Finally let us remark that, for $\beta\neq 1$, the conformal
mappings $f_r$ may be far more complicated than in the above example.
With $\beta$ rational they have a finite number of
branches that describe a Riemann surface. Like in the $\beta=1$ case,
only one of them will define the correct Neumann coefficients. When $\beta$
is irrational, as shown in (\ref{y_r}), this branch can always be determined,
but the Riemann surface interpretation is lost.

\section{The dTL Hirota equations}

Let us introduce the flow parameters $t^{(r)}_0, t^{(r)}_k, \bar t^{(r)}_k$
with $k= 1,2,..,\infty$ and the differential operators
\be
 D_r(z) = \sum_{k=1}^\infty \frac 1{kz^k} \frac {\d}{\d t^{(r)}_k},
\quad\quad \bar D_r(\bar z) = \sum_{k=1}^\infty \frac 1{k
{\bar z}^{~k}} \frac {\d }{\d {\bar t}^{~(r)}_k}, \quad r,s=1,2,3
\quad modulo \quad 3~.\label{DDbar}
\ee
The Hirota equations for the three decoupled copies of the dispersionless
Toda lattice hierarchies fit for the present case are
(for more details, see \cite{Zabrodin} and references therein)
%zdes
\be
&&(z_1-z_2) e^{D_r(z_1)D_r(z_2)F} = z_1 e^{-\d_{t^{(r)}_0}D_r(z_1)F}
- z_2 e^{-\d_{t^{(r)}_0}D_r(z_2)F},\label{HdTL1} \\
&&(\bar z_1-\bar z_2) e^{\bar D_r(\bar z_1)\bar D_r(\bar z_2)F} =
\bar z_1 e^{-\d_{t^{(r)}_0}\bar D_r(\bar z_1)F}
- \bar z_2 e^{-\d_{t^{(r)}_0}\bar D_r(\bar z_2)F},\label{HdTL2}\\
&& -z_1\bar z_2 \left( 1 - e^{-D_r(z_1)\bar D_r(\bar z_2)F}\right)=
e^{\d_{t^{(r)}_0}(\d_{t^{(r)}_0} + D_r(z_1)+\bar D_r(\bar z_2))F}
\label{HdTL3}
 \ee
where $F\equiv F(\{t^{(r)}_0, t^{(r)}_k, \bar t^{(r)}_k\})$ is
the $\tau$--function (free energy) of the hierarchy.
The minus sign on the l.h.s. of eq. (\ref{HdTL3}) can be replaced
by the standard plus sign via the transformations
$\{t^{r}_n, ~\bar t^{~r}_n\} \Rightarrow \{t^{r}_n, ~(-1)^n \bar t^{~r}_n\}$
or $\{t^{r}_n, ~\bar t^{~r}_n\}\Rightarrow \{(i)^n t^{r}_n,
~(i)^n \bar t^{~r}_n\}$.

We remark that eqs. (\ref{HdTL1}--\ref{HdTL3}) form three distinct sets
of equations, each one being
formally the same as in Witten's OSFT, \cite{BS}. The fact is that
in the latter case these sets of equations collapse to the same set.

Now, it is elementary to verify that the generating functions
$N^{r}(z), ~ N^{rs}(z_1, z_2)$ (\ref{Nr}--\ref{Nrs}) of the
Neumann coefficients satisfy the following equations:
 \be
(z_1-z_2) e^{N^{rr}(z_1,z_2)} &=& z_1 e^{-\beta_{r+1}\beta_r N^r(z_1)}
- z_2 e^{-\beta_{r+1}\beta_r N^r(z_2)}\0\\
&=& z_1 e^{\beta_{r}N^r(z_1)} - z_2 e^{\beta_{r}N^r(z_2)}
\label{HN1}
\ee
and
\be
-z_1 \bar z_2 \left( 1 - e^{N^{rr+1}(z_1, \bar z_2)}\right)
=e^{\tau_0(\frac{1}{\alpha_r}+ \frac{1}{\alpha_{r+1}})
+ \beta_{r+1}\beta_r N^r(z_1) - \beta_{r+1}N^{r+1}(\bar z_2)}.\label{HN2}
 \ee
These are easily seen to reproduce the Hirota equations (\ref{HdTL1}--\ref{HdTL3})
provided we suitably identify the generating functions with the second
derivatives of $F$ as follows:
\be
&&D_r(z_1)D_r(z_2) F = N^{rr}(z_1, z_2), \nonumber\\
&&\bar D_r(\bar z_1)\bar D_r(\bar z_2) F
= N^{r+1 r+1}(\bar z_1, \bar z_2), \nonumber\\
&& D_r(z_1)\bar D_r(\bar z_2) F =
- N^{r r+1}(z_1, \bar z_2),\nonumber\\
&&\d_{t^{(r)}_0} D_r(z) F = \beta_{r+1}\beta_r N^{r}(z),\nonumber\\
&&\d_{t^{(r)}_0} \bar D_r(\bar z) F =
-\beta_{r+1} N^{r+1}(\bar z),\nonumber\\
&&\d_{t^{(r)}_0}\d_{t^{(r)}_0}F
=\tau_0(\frac{1}{\alpha_r}+\frac{1}{\alpha_{r+1}})~.
\label{id1}
 \ee

However we remark that we can also define a consistent reduction by:
\be
\frac{\partial}{\partial \bar t^{~(r)}_n}F=\pm \frac{\partial}{\partial t^{(r+1)}_n}F.
\label{red1}
 \ee
This allows us to define new identifications as follows
 \be
&&D_r(z_1)D_r(z_2) F = N^{rr}(z_1, z_2), \nonumber\\
&& D_r(z_1)D_{r+1}(z_2) F = \mp N^{r r+1}(z_1, z_2),\nonumber\\
&&\d_{t^{(r)}_0} D_r(z) F = \beta_{r+1}\beta_r N^{r}(z),\nonumber\\
&&\d_{t^{(r)}_0} D_{r+1}(z) F =
\mp \beta_{r+1} N^{r+1}(z),\nonumber\\
&&\d_{t^{(r)}_0}\d_{t^{(r)}_0}F
=\tau_0(\frac{1}{\alpha_r}+\frac{1}{\alpha_{r+1}})~.
\label{id1-red1}
 \ee
The corresponding equations for the tau function are
\be
(z_1-z_2) e^{D_r(z_1)D_r(z_2)F}
&=& z_1 e^{-\d_{t^{(r)}_0}D_r(z_1)F}
- z_2 e^{-\d_{t^{(r)}_0}D_r(z_2)F}\0\\
&=& z_1 e^{\mp\d_{t^{(r-1)}_0}D_r(z_1)F}
- z_2 e^{\mp\d_{t^{(r-1)}_0}D_r(z_2)F},\0 \\
-z_1 z_2 \left( 1 - e^{\mp D_r(z_1)D_{r+1}(z_2)F}\right)
&=& e^{\d_{t^{(r)}_0}(\d_{t^{(r)}_0} + D_r(z_1)\pm D_{r+1}(z_2))F}.
\label{HdTL2-red1}
\ee
These Hirota equations refer to the dispersionless hierarchy
produced by the cyclic (coupled) reduction (\ref{red1}) of the three
copies of the dispersionless Toda lattice hierarchies. Our conjecture
is that the resulting hierarchy can be related to the 3-punctures Whitham
hierarchy
\cite{K} and the dispersionless limit of the 3-component KP hierarchy, but
the detailed analysis of this point is out of the scope of the present
paper and will be given elsewhere.

\section{Hirota equations for more general correlators}

As we will see in the next section, in order to verify the validity
of the full Hirota equations we need to know correlators with more than two
entries.
The dispersionless Hirota equations for `$n$--point functions' are obtained
by differentiating $n-2$ times the equations (\ref{HdTL1},\ref{HdTL2}) and
(\ref{HdTL3}). These derived Hirota equations, like the original ones,
do not uniquely determine their solutions. In order to be able to write
down the latter we have to provide some additional information. In fact
what we are looking for are $n$--point correlators that are compatible
with the Neumann coefficients of the previous section and satisfy
the full Hirota equation (see next section). Although we don't have a proof
of it, we believe these two requirements completely determine the full
series of $n$--point functions.

In this section we would like to concentrate
on three-- and four--point functions. On the basis of what we have just said,
in order to determine them we have to rely on plausibility arguments and
verify the results {\it a posteriori}. We notice first that
the Neumann coefficients for the three strings vertex can be interpreted
in terms of two--point correlators of a 1D system. For instance
\be
\langle V_3| a_m^{(r)I\dagger} a_n^{(s)J\dagger}|0\rangle
=\delta^{IJ}\, V_{mn}^{rs}=
- \sqrt{nm}\, N^{rs}_{mn}\,\delta^{IJ}~.\label{2pt}
\ee
Analogous relations hold for two--point functions involving zero modes.
In the same way we can of course consider correlators with more insertions.
It is obvious that all the correlators with an odd number of insertions
identically vanish. We take these correlators as a model in order to calculate
three-- and four--point functions of a quantum system that satisfies
the full Hirota equations. We will see below that the correct answer
is not the three-- and four--point correlators obtained by inserting one and
two creation operators in (\ref{2pt}), respectively, but combinations
of them. We will refer to the underlying model, based on the two--point
correlators (\ref{2pt}) and satisfying the Hirota equation, as the
{\it associated quantum system}. It must be clear that we do not know yet
what this system is, we simply postulate its existence. In particular,
as a working hypothesis, we assume that its
three--point functions identically vanish.

The dispersionless Hirota equations for four--point functions are obtained
by differentiating twice the equations (\ref{HdTL1},\ref{HdTL2}) and
(\ref{HdTL3}). Of course we have the possibility of applying $\d_{t_0^{(r)}}$
or $D_r(z)$. Therefore from (\ref{HdTL1}), for instance, we obtain
\be
&&(z_1-z_2) e^{D_r(z_1)D_r(z_2)F} \prod_{i=1}^{4}D_r(z_i)F=
\label{HdTL11}\\
&&~~~~~z_2 e^{-\d_{t^{(r)}_0}D_r(z_2)F}\d_{t^{(r)}_0}D_r(z_2)\prod_{i=3}^{4}D_r(z_i)F-z_1
e^{-\d_{t^{(r)}_0}D_r(z_1)F}\d_{t^{(r)}_0}D_r(z_1)\prod_{i=3}^{4}D_r(z_i)F,\0\\
&&(z_1-z_2) e^{D_r(z_1)D_r(z_2)F} \d_{t^{(r)}_0} \prod_{i=1}^{3}D_r(z_i)F=
\label{HdTL12}\\
&&~~~~~~z_2 e^{-\d_{t^{(r)}_0}D_r(z_2)F}\d_{t^{(r)}_0}^2D_r(z_2)D_r(z_3) F-z_1
e^{-\d_{t^{(r)}_0}D_r(z_1)F}\d_{t^{(r)}_0}^2D_r(z_1)D_r(z_3) F,\0\\
&&(z_1-z_2) e^{D_r(z_1)D_r(z_2)F} \d_{t^{(r)}_0}^2D_r(z_2)D_r(z_1)F=
\label{HdTL13}\\
&&~~~~~~z_2 e^{-\d_{t^{(r)}_0}D_r(z_2)F}\d_{t^{(r)}_0}^3D_r(z_2)  F-z_1
e^{-\d_{t^{(r)}_0}D_r(z_1)F}\d_{t^{(r)}_0}^3D_r(z_1)F\0
\ee
where we have assumed that three--point correlators vanish.

After expanding the above equations in powers of $z_1$ and $z_2$, and
collecting terms of the form $z_1^m z_2^n$ we obtain equations involving
two and four derivative of $F$. We already know how
to identify the two--derivative terms in terms of the Neumann coefficients
of LCSFT (see previous section). The task now is to
try to make the corresponding identifications for the four--derivative ones.
The solutions to eqs. (\ref{HdTL11}--\ref{HdTL13}) are not uniquely
defined (see below). The solutions we are interested in are as follows:
\be
&&F_{t_0^{(r)}t^{(r)}_0t^{(r)}_0t^{(r)}_m}=-3(\alpha_r)^2\frac{1+\beta_r}
{\beta_r}m(N_{m0}^{rr+1}-N_{m0}^{rr}),\nonumber\\
&&F_{t_0^{(r)}t^{(r)}_0t^{(r)}_mt^{(r)}_n}=
-(\alpha_r)^2mn\left[\frac{1+\beta_r}{\beta_r}(n+m)N^{rr}_{mn}+
2(N_{m0}^{rr+1}-N_{m0}^{rr})(N_{n0}^{rr+1}-N_{n0}^{rr})\right],\nonumber\\
&&F_{t_0^{(r)}t^{(r)}_mt^{(r)}_nt^{(r)}_l}=-(\alpha_r)^2mnl[(n+l)(N_{m0}^{rr+1}-
N_{m0}^{rr})N^{rr}_{nl}+(n+m)(N_{l0}^{rr+1}-N_{l0}^{rr})N^{rr}_{mn}\nonumber\\
&&~~~~~~~~~~~~~~~~~~~+(m+l)(N_{n0}^{rr+1}-N_{n0}^{rr})N^{rr}_{ml}],
\label{identific}\\
&&F_{t_m^{(r)}t^{(r)}_nt^{(r)}_lt^{(r)}_k}=
-(\alpha_r)^2mnlk[(m+n)(l+k)N^{rr}_{mn}N^{rr}_{lk}+
(m+l)(n+k)N^{rr}_{ml}N^{rr}_{nk}\nonumber\\
&&~~~~~~~~~~~~~~~~~~~+(m+k)(n+l)N^{rr}_{mk}N^{rr}_{nl}]\0
\ee
where $N_{m0}^{rs}$ are defined in eqs. (\ref{VN}).
Multiplying these by the appropriate monomials of $z$ variables and summing,
leads to the following compact expressions (which will be used
later on):
\be
\partial^3_{t^{(r)}_0}D_r(z)F
&=&\frac{3 \a_{r-1}^2}{\beta_r}\,
\frac{\varphi_r(z^{-1})-1}{(1+\beta_r)\varphi_r(z^{-1})-\beta_r},
\label{4point} \\
\partial^2_{t^{(r)}_0}\prod_{i=1}^{2}D_r(z_i)F&=&
-3 \a_{r-1}^2\, \prod_{i=1}^2
\frac{\varphi_r(z_i^{-1})-1}{(1+\beta_r)\varphi_r(z_i^{-1})- \beta_r}\0 ,
\nonumber\\
\partial_{t^{(r)}_0}\prod_{i=1}^{3}D_r(z_i)F&=&
3\alpha_{r-1}^2\beta_r\,
\prod_{i=1}^3
\frac{\varphi_r(z_i^{-1})-1}{(1+\beta_r)\varphi_r(z_i^{-1})- \beta_r}\0 ,
\nonumber\\
\prod_{i=1}^{4}D_r(z_i)F&=&-3\alpha_{r-1}^2\beta_r^2\,
\prod_{i=1}^4
\frac{\varphi_r(z_i^{-1})-1}{(1+\beta_r)\varphi_r(z_i^{-1})- \beta_r}\0 .
\ee
Using these expressions one can easily verify that they satisfy equations
(\ref{HdTL11}), (\ref{HdTL12}) and (\ref{HdTL13}).

A comment is in order concerning the results we have just written down.
We have already said that eqs. (\ref{4point}) are not unique solutions to
the Hirota equations for four--point functions. We have single them out
because they are compatible with the dispersive Hirota equations
(see next section) and they will allow us to find a solution of the latter
coherent with the dispersionless Neumann coefficients.
We recall that they are also compatible with vanishing three--point functions.
As was mentioned at the beginning of this section, at first sight it would
seem that the two--point functions,
three--point functions and four--point functions (i.e. the derivatives of $F$
with respect to two, three and four $t_n$ parameters) of our system are
simply given
by
\be
\langle V_3|a_{n_1}^{r_1\dagger}\ldots a_{n_k}^{r_k\dagger}|0\rangle .
\label{kpoint}
\ee
While this is true for two-- and three--point functions, it is not quite
true for the four--point ones, as one can see by comparing (\ref{kpoint}) with
(\ref{identific}). It is evident that the four--point functions given by
(\ref{kpoint}) has the right form, but must be suitably combined
in order to coincide with (\ref{identific}), which satisfy the Hirota equations.
Therefore the correlators of the associated quantum system that underlies the Neumann
coefficients of LCSFT are made of suitable combinations of the
$k$--point functions (\ref{kpoint}). In conclusion: we have found a solution
to the Hirota equations for the four--point functions, which is compatible with
the Mandelstam's three strings vertex of the LCSFT (and with its deformations
up to second order in the expansion parameter, see next section), however
we do not have a proof that this solution is unique.

A discussion of the general solution to
eqs. (\ref{HdTL11}--\ref{HdTL13}) can be found in Appendix A.

\section{PP--wave SFT and the dispersive Hirota equations}

String theory on a maximally supersymmetric pp--wave background
(plane wave limit of $AdS_5\times S^5$)
\cite{Blau,BMN}, is exactly solvable \cite{metsaev}. Building on this it has been
recently possible to construct the exact three strings vertex for the LCSFT
on this background, i.e. to completely specify the relevant Neumann
coefficients. They depend on the `string mass' $\mu$ (determined by the
five form flux of type IIB superstring theory). When $\mu\to 0$ one recovers
Mandelstam's Neumann coefficients discussed in the previous sections.
In other words the nontrivial string background deforms the Neumann
coefficients. It is interesting to see whether the Hirota equations discussed
in the previous sections get deformed accordingly in such a way as to preserve
integrability. This is our guess and this is what we would like to provide
evidence for in the remaining part of our paper.

Let us start from the expression of the $\mu$--deformed three strings vertex,
\cite{Sprad1,Huang,Chu,Sprad2,Schwarz,Pank,He,Lucietti1,Lucietti2,Lucietti3}.
The bosonic part, to which we limit ourselves\footnote{For the problems 
connected with the supersymmetric completion of the vertex and its prefactor see
the reviews \cite{Maldacena,Sadri,Russo} and references therein.}, is defined by
\be
 \Delta'_B=\frac 12 \delta_{IJ}\left(
\sum_{r,s=1}^3 \sum_{m,n=-\infty}^{\infty} a_m^{(r)I\dagger}
\EV_{mn}^{rs}a_n^{(s)J\dagger} \right)~.\label{Vmu}
\ee
The commutation relations are as in section 2, but $a_m^{(r)I\dagger}\neq
a_{-m}^{(r)I}$ and $a_m^{(r)I}|0\rangle=0$ for $m\in {\mathbb Z}$. The
coefficients with negative $n,m$ labels can be obtained from those with
positive ones\footnote{There is a subtle point here.
The expressions of $\Delta'_B$ and $\Delta_B$ in
eq.(\ref{DeltaB}) for the three strings vertex refer to two different
bases: $\Delta_B$ is expressed in terms of the transverse momenta $p^I_r$
(momentum basis) while $\Delta'_B$ contains
the zero mode oscillators $a_0^{(r)I}$ (oscillator basis). This means that
starting from (\ref{V3}) we have passed from the former basis to the latter
by explicitly integrating over the transverse momenta (see, for
instance, \cite{RSZ}). This operation modifies in a well--known way the
vertex coefficients. However, in the present case, it turns out that
in the limit $\mu\to 0$ the coefficients $\EV_{mn}^{rs}$ tend to the
corresponding coefficients $V_{mn}^{rs}$ in (\ref{DeltaB}) at least
for $m,n\geq 1$.}. In this paper we consider only the $\EV_{mn}^{rs}$ with
$m,n\geq 0$.
We set
\be
\EV_{mn}^{rs} = -\sqrt{mn}\, \EN_{mn}^{rs}~.\label{EVN}
\ee
The Neumann coefficients $ \EN_{mn}^{rs}$ with $m,n\geq 1$ have been calculated
in an explicit way in \cite{Lucietti2,Lucietti3}:
\be
\EN_{mn}^{rs} = - \frac {mn \a} {\a_s \omega_{r,m}+\a_r \omega_{s,n}} \,
\frac{\EN_m^r \EN_n^s}{\a_r\a_s}\label{ENnm}
 \ee
where $\omega_{r,m}= \sqrt{m^2 +\a_r^2\mu^2}$ and
\be
\EN_m^r= \sqrt{\frac {\omega_{r,m}}{m}}\,
\frac{\omega_{r,m}+\a_r\mu}{m} f_{m}^{(r)}~.\label{ENm}
\ee
In these formulas
\be
f_{m}^{(r)}= - \frac{e^{\tau_0 \omega_{\frac m{\a_r}}}}
{m(\a_r+\a_{r+1})\omega_{\frac m{\a_r}}} \, \frac{\Gamma_\mu^{(r+1)}
\left(-\frac m{\a_r}\right)}  {\Gamma_\mu^{(r)}\left(\frac m{\a_r}\right)
\Gamma_\mu^{(r-1)}\left(\frac m{\a_r}\right)}
\ee
and $\omega_z={\rm sgn}(z)\sqrt {\mu^2+z^2}$.
The $\mu$--deformed $\Gamma$ functions are defined by
\be
\Gamma_\mu^{(r)}(z) = \frac{e^{-\gamma \a_r \omega_z}}{\a_r z}
\prod_{n=1}^\infty \left(\frac n{\omega_{r,n}+\a_r
\omega_z}e^{\frac{\a_r\omega_z}n}\right)\label{Gammadef}
\ee
where $\gamma$ is the Euler--Mascheroni constant. It must be remarked that
the above coefficients are not written in the usual form, we have simplified
them by dropping some intermediate inessential factors.

Here we present some results which are going to be needed in the next sections.
Expanding (\ref{ENm}) and (\ref{ENnm}) in powers of $\mu$ up to order $\mu^2$
we obtain
\be
\EN_m^r=N_m^r\left[1+{\alpha_{r}\over m}\mu+\left({\alpha_{r}^2\over 4m^2}+
{\alpha_{r}\over 2m}\tau_{0}\right)\mu^2+...\right]\label{ENr1}
\ee
and
\be
\EN_{mn}^{rs}=N_{mn}^{rs}\left[1+\left({\alpha_{r}\over m}+
{\alpha_{s}\over n}\right)\mu+{1\over 4}\left({\alpha_{r}\over m}+
{\alpha_{s}\over n}\right)\left({\alpha_{r}\over m}+{\alpha_{s}\over n}+
2\tau_{0}\right)\mu^2+...\right]\label{ENrs1}
\ee
where $N_m^r$ and $N_{mn}^{rs}$ are Mandelstam's Neumann coefficients.
For reasons that will become clear in the next section, it is convenient
to rewrite these expansions in powers of $\lambda$ related to $\mu$ as:
$\mu=\lambda-{\tau_{0}\over 2}\lambda^2+...$
\be
\EN_m^r=N_m^r\left[1+{\alpha_{r}\over m}\lambda+
\left({\alpha_{r}^2\over 4m^2}\right)\lambda^2+...\right]\equiv
\sum_{j=0}^{\infty}\EN_{j,m}^r\lambda^j,\label{ENr2}
\ee
\be
\EN_{mn}^{rs}=N_{mn}^{rs}\left[1+\left({\alpha_{r}\over m}+
{\alpha_{s}\over n}\right)\lambda+{1\over 4}\left({\alpha_{r}\over m}+
{\alpha_{s}\over n}\right)^2\lambda^2+...\right]\equiv
\sum_{j=0}^{\infty}\EN_{j,mn}^{rs}\lambda^j\label{ENrs2}
\ee
where the subscript $j$ refers to the order of expansion in $\lambda$.

Again, for later use, it is convenient to organize these Neumann coefficients
by means of associated generating functions
\be
\EN_j^{r}(z):=\sum_{m=1}^{\infty}\frac{1}{z^m}~m^{2j}
~\EN_{j,m}^r, \label{Nir}
\ee
\be
\EN_j^{rs}(z_1,z_2):=\sum_{m,n=1}^{\infty}
\frac{1}{z_1^m}\frac{1}{z_2^n}~(mn)^j~\EN_{j,mn}^{rs} ~,
\label{Nirs}
\ee
for $j=0,1,2$. At each order the summation can easily be carried out to give
\be
\EN_1^r(z)=\alpha_{r}\frac{\varphi_r(z^{-1})-1}
{(1+\beta_r)\varphi_r(z^{-1})-\beta_r},
\ee
\be
\EN_2^r(z)=\frac{\alpha^2_r}{4}{\varphi_r(z^{-1})(\varphi_r(z^{-1})-1)
\over ((1+\beta_r)\varphi_r(z^{-1})-\beta_r)^3},
\ee
\be
&&\EN_1^{rs}(z_1,z_2)=-{\alpha\over \alpha_{r}\alpha_{s}}
~{\varphi_r(z_1^{-1})-1
\over (1+\beta_r)\varphi_r(z_1^{-1})-\beta_r}~{\varphi_s(z_2^{-1})-
1\over (1+\beta_s)\varphi_s(z_2^{-1})-\beta_s},\label{N1rs}\\
&&\EN_2^{rs}(z_1,z_2)=-{\alpha\over 4\alpha_{r}\alpha_{s}}
\frac{(\varphi_r(z_1^{-1})-1)(\varphi_s(z_2^{-1})-1)}
{[(1+\beta_r)\varphi_r(z_1^{-1})-\beta_r][(1+\beta_s)\varphi_s(z_2^{-1})-
\beta_s]}\0\\
&&~~~~~~~~~~\times \left[\frac{\alpha_s \varphi_r(z_1^{-1})}
{[(1+\beta_r)\varphi_r(z_1^{-1})-\beta_r]^2}+
\frac{\alpha_r \varphi_s(z_2^{-1})}{[(1+\beta_s)\varphi_s(z_2^{-1})-
\beta_s]^2}\right]~.\label{N2rs}
\ee

We notice that in order to get these compact generating functions it is
necessary to insert the $m^{2j}$ and $(mn)^{j}$ factors in (\ref{Nir})
and (\ref{Nirs}), respectively.

\subsection{The dispersive Hirota equations}

Here, we consider the full (dispersive) Hirota equations for the Toda lattice
hierarchy (see, e.g. ref. \cite{Zabrodin} and references therein). For the
sake of brevity we deal in the sequel with half of the story, namely only
with those equations that involve unbarred $z$ variables (corresponding to
eq.(\ref{HdTL1}))
\be
&&z_1 \left(e^{\lambda \left(\partial_{t_0}-D_r(z_1)\right)}
\tau_\lambda\right) \left(e^{-\lambda D_r(z_2)}\tau_\lambda\right)-
z_2 \left(e^{\lambda \left(\partial_{t_0}-D_r(z_2)\right)}
\tau_\lambda\right) \left(e^{-\lambda D_r(z_1)}\tau_\lambda\right)\label{qHE}\\
&&= (z_1-z_2)\left(e^{-\lambda \left(D_r(z_1)+D_r(z_2)\right)}
\tau_\lambda\right)\left(e^{\lambda\partial_{t_0} }\tau_\lambda\right)
\0
\ee
where $\lambda$ is a deformation parameter and $\tau_\lambda$ is the full
{\it tau function} of the Toda lattice (KP) hierarchy:
\be
\tau_\lambda = {\rm exp}({\EF_\lambda}),\quad\quad
\EF_\lambda= \frac 1{\lambda^2} F_0 + \frac 1{\lambda} F_1+
F_2+\ldots \quad .\label{tauh}
\ee
In order to find  a solution to (\ref{qHE}) it is useful to proceed in
two steps, and solve first the Hirota equation that does not involve
$t_0$ derivatives (corresponding to the KP hierarchy), that is
\be
(z_1-z_2)\left(e^{-\lambda \left(D_r(z_1)+D_r(z_2)\right)}
\tau_\lambda\right)\left(e^{-\lambda
D(z_3)}\tau_\lambda\right)+ {\rm cycl.\, perms.\, of}\quad 1,2,3=0.\label{qHEs}
\ee

Our conjecture is that
these equations are obeyed by the Neumann coefficients $\EN_{mn}^{rs}$
introduced above, provided we identify $\lambda$ with a suitable function
$f(\mu)$ of $\mu$. At present we are not in the condition to prove this
conjecture in a
non--perturbative way. But we can expand in powers of $\lambda$ and try
to prove it order by order in $\lambda$. We will be able to do it up to second
order in $\lambda$, with the identification
$\lambda = \mu +\frac {\tau_0}2 \mu^2+\ldots$~.

Expanding (\ref{qHE}) in powers of $\lambda$ we obtain an infinite set of
equations that constrain the correlators at the different orders of
approximation. To order 0 we get the dispersionless Hirota equation.
Our task is therefore to prove that the next order equations hold.

\subsection{Order $\mu$ approximation}

Expanding (\ref{qHEs}) to first order in $\lambda$ and identifying
$\lambda$ with $\mu$ we find
\be
(z_1-z_2) e^{D_r(z_1)D_r(z_2)F_0} D_r(z_1)D_r(z_2)F_1 +
{\rm cycl.\, perms.\, of}\quad 1,2,3=0 ~ \label{qHE1st}
\ee
while, expanding (\ref{qHE}), we get
\be
-z_1e^{-\partial_{t_0^{(r)}}D_r(z_1)F_0}\partial_{t_0^{(r)}}D_r(z_1)F_1+
z_2e^{-\partial_{t_0^{(r)}}D_r(z_2)F_0}\partial_{t_0^{(r)}}D_r(z_2)F_1 \0 \\
~~~~~~~~~~~~~~~~~=(z_1-z_2) e^{D_r(z_1)D_r(z_2)F_0} D_r(z_1)D_r(z_2)F_1.
\label{qHE1stom}
\ee
In deriving these equations we have used the information that all third order
derivatives of $F_0$ (three--point functions) vanish (see previous section).
It is easy to see that
if we make the following identifications:
\be
D_r(z_1)D_r(z_2)F_1= \EN_1^{rr}(z_1,z_2)\label{Dz1Dz2F1}
\ee
with $D_r(z_1)D_r(z_2)F_0$ as in (\ref{id1}), eq.(\ref{qHE1st}) is satisfied.
If in addition we identify
\be
\partial_{t_0^{(r)}}D_r(z)F_1=\beta_{r+1} \beta_{r}\EN_1^{r}(z),\label{dt0DzF1}
\ee
with $\partial_{t_0^{(r)}}D_r(z)F_0$ as in
(\ref{id1}), the more general eq.(\ref{qHE1stom}) is satisfied as well.

\subsection{Order $\mu^2$ approximation}

Expanding (\ref{qHEs}) to order $\lambda^2$ and identifying
$\lambda$ with $\mu +\frac {\tau_0}2 \mu^2+\ldots$ we find
\be
&&(z_1-z_2) e^{D_r(z_1)D_r(z_2)F_0} \left(D_r(z_1)D_r(z_2)F_2
+\frac 12 \left(D_r(z_1)D_r(z_2)F_1\right)^2 +
 \frac 16 D_r(z_1)D_r(z_2)^3F_0\right.\0\\
&&~~~~~~~+\left.\frac 14 D_r(z_1)^2 D_r(z_2)^2F_0+
\frac 16 D_r(z_1)^3D_r(z_2)F_0\right)
+{\rm cycl.\, perms.\, of}\, 1,2,3=0 ~ \label{qHE2nd}
\ee
while expanding (\ref{qHE}) to the same order we get
\be
z_1e^{-\partial_{t_0^{(r)}}D_r(z_1)F_0}\left(- \partial_{t_0^{(r)}}D_r(z_1)F_2
+{1\over 2}(\partial_{t_0^{(r)}}D_r(z_1)F_1)^2+
{1\over 4}\partial_{t_0^{(r)}}^2D_r(z_1)^2 F_0-
{1\over 6}\partial_{t_0^{(r)}}D_r(z_1)^3 F_0\right.\0\\
\left.-{1\over 6}\partial_{t_0^{(r)}}^3D_r(z_1)F_0\right)
-z_2e^{-\partial_{t_0^{(r)}}D_r(z_2)F_0}\left(- \partial_{t_0^{(r)}}D_r(z_2)F_2
+{1\over 2}(\partial_{t_0^{(r)}}D_r(z_2)F_1)^2\right.\0\\
\left.+{1\over 4}\partial_{t_0^{(r)}}^2D_r(z_2)^2 F_0-
{1\over 6}\partial_{t_0^{(r)}}D_r(z_2)^3 F_0-
{1\over 6}\partial_{t_0^{(r)}}^3D_r(z_2)F_0\right)\0 \\
=(z_1-z_2) e^{D_r(z_1)D_r(z_2)F_0} \left(D_r(z_1)D_r(z_2)F_2
+\frac 12 \left(D_r(z_1)D_r(z_2)F_1\right)^2 +
 \frac 16 D_r(z_1)D_r(z_2)^3F_0\right.\0 \\
+\left.\frac 14 D_r(z_1)^2 D_r(z_2)^2F_0+
\frac 16 D_r(z_1)^3D_r(z_2)F_0\right).~~~~~~~~~~~~~~ \label{qHE2ndom}
\ee
In deriving this equation we have used once again the information that all
odd order
derivatives of $F_0$ and $F_1$ vanish.
Eq.(\ref{qHE2nd}) is satisfied provided we make the following identifications:
\be
D_r(z_1)D_r(z_2)F_2=\EN_2^{rr}(z_1,z_2)
\label{Dz1Dz2F2}
\ee
with four-derivatives
as defined in (\ref{4point}). If, in addition, we identify
\be
\partial_{t_0^{(r)}}D_r(z)F_2=\beta_{r+1} \beta_{r}\EN_2^{r}(z)
\label{dt0DzF2}
\ee
the more general equation (\ref{qHE2ndom}) is satisfied.

Taking into account eqs. (\ref{id1}), (\ref{Dz1Dz2F1}--\ref{dt0DzF1})
and (\ref{Dz1Dz2F2}--\ref{dt0DzF2}) it is plausible to expect that
the following universal identification is valid in general:
\be
D_r(z_1)D_r(z_2)F_j=\EN_j^{rr}(z_1,z_2), \quad
\partial_{t_0^{(r)}}D_r(z)F_j=\beta_{r+1} \beta_{r}\EN_j^{r}(z), 
\quad j=0,1,2,3,...~
\label{ident-gen}
\ee
although in this
case one may have to modify the definitions (67-68) of the
generating functions for $j > 2$.

\section{Discussion}

In this paper, to start with, we have shown that the three strings vertex
coefficients (or, more properly, the corresponding Neumann coefficients) in
LCSFT obey the Hirota equations for the dispersionless Toda lattice hierarchy.
We have then written down the Hirota equations for the four--point functions
and found solutions consistent with identically vanishing three--point functions.
Finally we have expanded the three strings vertex of the LCSFT in a maximally
supersymmetric pp--wave background in a series in the $\mu$ parameter, and,
in a parallel way, we have expanded the Hirota equations for the full Toda lattice
hierarchy in the deformation parameter. After a (non--trivial) identification
of the two parameters we have been able to show that the latter are satisfied
by the former up to second order. The calculations beyond second order become
technically very challenging and it is clear that a non--perturbative approach
is needed for a satisfactory proof that LCSFT three--strings vertex
obey the Hirota equations for the full Toda lattice hierarchy.

We would like to end this paper with a few open questions. The first concerns
the type of hierarchy underlying the three strings vertex of LCSFT. We have
always mentioned the Toda lattice hierarchy, but one should be more precise.
Looking at section 3 it is evident that one has to do not with one Toda lattice
hierarchy but rather with three coupled copies of the Toda lattice hierarchy.
As we mentioned in the same section, there is the possibility of
a consistent reduction. A precise formulation and identification of the
relevant reduced hierarchies is a task we have not tackled in this paper
(see, however, the remark at the end of section 3).

As we have pointed out in section 4, the four--point functions we have
found are not the unique solutions to the Hirota equations for
four--point functions. The one that we have found however are likely to be the
unique solution that are consistent with the two--point functions of order
0, 1 and 2 in the $\mu$ expansion  (although we have not been able to exclude
other solutions). This question is intertwined with two related problems:
on the one hand the question of defining in terms of matter oscillators $a_n^{(r)}$
the 1D associated quantum system whose two--point correlators are given by $\langle V_3|
a_m^{(r)\dagger}a_n^{(s)\dagger}|0\rangle$ and underlies
the LCSFT three strings vertex\footnote{We have seen in
section 4 that the four--point functions that are required by the Hirota
equations are not obtained by simple adjunction of two creation operators
in this expression (see the final remark there).}; on the other hand
defining an integrable
system, very likely a matrix model, where all these correlators can
explicitly be calculated. A successful search of such an integrable model
is also likely to lead us to a natural explanation of why the somewhat
mysterious factors $m^{2j}$ and $(nm)^j$ need to be inserted in the
generating functions (\ref{Nir}--\ref{Nirs}) in order to square matters.

~{}

~{}

{\bf Acknowledgments}
We would like to thank L. Martinez Alonso, B. Dubrovin, P. Fr\'e, B. Konopelchenko,
O. Lechtenfeld, I. Loutsenko, M. Ma${\tilde {\rm n}}$as, K. Takasaki and
A.Tanzini for useful
discussions. A.S. would like to thank SISSA for the kind hospitality during
the course of this work. This research was partially supported by the Italian
MIUR under the program ``Teoria dei Campi, Superstringhe e Gravit\`a'',
by the RFBR-DFG Grant No. 04-02-04002, DFG Grant 436 RUS 113/669-2, the NATO Grant
PST.GLG.980302, by the Heisenberg-Landau program, and by CAPES-Brasil as RJSS is concerned.

%\newpage

\subsection*{Appendix A. A discussion of four--point functions}
Using Hirota equations (\ref{HdTL11}--\ref{HdTL13}) one can easily
express the four--point functions \\ $\prod_{i=1}^{4}D_r(z_i)F$,
$\d_{t^{(r)}_0} \prod_{i=1}^{3}D_r(z_i)F$ and
$\d_{t^{(r)}_0}^2\prod_{i=1}^{2}D_r(z_i)F$
in terms of the two--point functions $\prod_{i=1}^{2}D_r(z_i)F$ and
$\d_{t^{(r)}_0}D_r(z)F$
as well as the four--point function $\d_{t^{(r)}_0}^3D_r(z)F$.
The two--point functions
were already identified with the corresponding generating functions
of the Neumann coefficients (see, eqs. (\ref{id1})), so one is left
with the problem of first finding a suitable identification for
the four--point function
$\d_{t^{(r)}_0}^3D_r(z)F$, which leads to the straightforward derivation
of all other.
It is interesting that if one additionally requires that the four--point
functions admit a factorizable form as functions of $z_i$, then the
identification of $\d_{t^{(r)}_0}^3D_r(z)F$ is uniquely fixed modulo
three arbitrary $SL(2)$-group parameters $a_r,~b_r,~ c_r$ and $d_r$
$(b_rc_r-a_rd_r=1)$ as follows
\be
\d_{t^{(r)}_0}^3D_r(z)F=
\frac{\varphi_r(z^{-1})-1}{\varphi_r(z^{-1})}~
\frac{a_r \varphi_r(z^{-1})+b_r}{c_r\varphi_r(z^{-1})+d_r}~.
\label{four-point1}
 \ee
Using this and eqs. (\ref{HdTL11}--\ref{HdTL13}), one can derive the
corresponding explicit expressions for the remaining four--point functions
\be
\d_{t^{(r)}_0}^2\prod_{i=1}^{2}D_r(z_i)F&=&
-\prod_{i=1}^{2} \frac{\varphi_r(z_i^{-1})-1}{c_r\varphi_r(z_i^{-1})+d_r}~,
\label{four-point2} \\
\d_{t^{(r)}_0} \prod_{i=1}^{3}D_r(z_i)F&=&
-d_r\prod_{i=1}^{3} \frac{\varphi_r(z_i^{-1})-1}{c_r\varphi_r(z_i^{-1})+d_r}~,
\label{four-point3}\\
\prod_{i=1}^{4}D_r(z_i)F&=&
-d^2_r\prod_{i=1}^{4} \frac{\varphi_r(z_i^{-1})-1}{c_r\varphi_r(z_i^{-1})+d_r}~.
\label{four-point4}
 \ee
The four--point functions (\ref{4point}) are a particular case of the latter
for
\be
d_r= - \frac{1}{a_r}=-\frac{\beta_r}{\sqrt{3}~\a_{r-1}},
\quad b_r=0, \quad c_r= -\frac{1}{\sqrt{3}~\a_r}.
\ee

Let us remark that if one rescales the parameters
in eqs. (\ref{four-point1}--\ref{four-point4}) as follows
\be
\{a_r,~b_r, ~c_r, ~d_r\} \quad \Rightarrow \quad
\frac{1}{\epsilon}~\{a_r,~b_r, ~c_r, ~d_r\}
\ee
and consider the limit $\epsilon \rightarrow 0$, then the four--point
function $\d_{t^{(r)}_0}^3D_r(z)F$ preserves its form (\ref{four-point1})
with the
parameters satisfying $b_rc_r-a_rd_r=0$, so it becomes
\be
\d_{t^{(r)}_0}^3D_r(z)F=
\frac{\varphi_r(z^{-1})-1}{\varphi_r(z^{-1})}~
\frac{a_r}{c_r}~,
\label{four-point5}
 \ee
but all the other four--point
functions degenerate and become equal to zero,
$\d_{t^{(r)}_0}^2\prod_{i=1}^{2}D_r(z_i)F=\d_{t^{(r)}_0}
\prod_{i=1}^{3}D_r(z_i)F=
\prod_{i=1}^{4}D_r(z_i)F=0$.

%\newpage


\begin{thebibliography}{99}

\bibitem{BS} L.~Bonora and A.~S.~Sorin,
  {\it Integrable structures in string field theory,}
  Phys.\ Lett.\ B {\bf 553} (2003) 317,
  [arXiv:hep-th/0211283].
  %%CITATION = HEP-TH 0211283;%%

\bibitem{W1} E. Witten, {\it Noncommutative Geometry and String Field Theory},
Nucl.Phys. {\bf B268} (1986) 253.

\bibitem{TL} R.~Hirota, J.Phys.Soc.Japan {\bf 50} (1981) 3785.

\bibitem{TT} K. Takasaki and T. Takebe, {\it Integrable hierarchies and
dispersionless limit}, {\it Rev.Math.Phys.} {\bf 7} (1995) 743,
[arXiv:hep-th/9405096].

\bibitem{CD} R. Carroll and Y. Kodama, {\it Solution of the
dispersionless Hirota equations}, J.Phys. {\bf A28} (1995) 6373,
[arXiv:hep-th/9506007].

\bibitem{Zabrodin} A. Zabrodin, {\it Dispersionless limit of
Hirota equations in some problems of complex analysis},
Theor.Math.Phys. {\bf 129} (2001) 1511,
(Teor.Mat.Fiz. {\bf 129} (2001) 239), [arXiv:math.CV/0104169].

\bibitem{LPPI} A. Le Clair, M.E. Peskin and C.R. Preitschopf, {\it
String Field Theory on the conformal plane (I). Kinematical
Principles}, Nucl.Phys. {\bf B317} (1989) 411.

\bibitem{BR} A.~Boyarsky and O.~Ruchayskiy,
  {\it Integrability in SFT and new representation of KP tau--function}
  JHEP {\bf 0303} (2003) 027,
  [arXiv:hep-th/0211010].
  %%CITATION = HEP-TH 0211010;%%

A.~Boyarsky, B.~Kulik, O.~Ruchayskiy, {\it  String Field Theory Vertices,
Integrability and Boundary States}, JHEP 0311 (2003) 045,
[arXiv:hep-th/0307057].

\bibitem{UT} K. Ueno and K. Takasaki, {\it Toda lattice hierarchy}, {\it Adv.
Stud. in Pure Math.}  {\bf 4} (1984)

\bibitem{Mandel} S.~Mandelstam, {\it Interacting--string picture of dual
resonance models.} Nucl.Phys. {\bf B64} (1973) 205.

\bibitem{GS}  M.~B.~Green and J.~H.~Schwarz, {\it Superstring interactions},
Nucl.Phys. {\bf B218} (1983) 43;
{\it Superstring field theory}, Nucl.Phys. {\bf B243} (1984) 475.

\bibitem{GSW}  M.~B.~Green, J.~H.~Schwarz and E.~Witten,
{\it Superstring theory: 2}, Cambridge University Press, Cambridge, 1987.

\bibitem{Kaku} M.Kaku, {\it Introduction to superstrings}, Springer--Verlag,
1988.

\bibitem{K} I. Krichever, {\it The tau function of the universal Whitham
hierarchy, matrix models and topological field theories},
Commun.Pure.Appl.Math. {\bf 47} (1992) 437, [arXiv:hep-th/9205110].

\bibitem{Blau} M.~Blau, J.~Figueroa-O'Farrill, C.~Hull and G.~Papadopoulos,
  {\it A new maximally supersymmetric background of IIB superstring theory,}
  JHEP {\bf 0201} (2002) 047,
  [arXiv:hep-th/0110242].

\bibitem{BMN} D.~Berenstein, J.~M.~Maldacena and H.~Nastase,
  {\it Strings in flat space and pp waves from N = 4 super Yang Mills,}
  JHEP {\bf 0204} (2002) 013,
  [arXiv:hep-th/0202021].
  %%CITATION = HEP-TH 0202021;%%

\bibitem{metsaev} R.~R.~Metsaev,
  {\it Type IIB Green-Schwarz superstring in plane wave Ramond-Ramond
  background,}
  Nucl.\ Phys.\ B {\bf 625} (2002) 70,
  [arXiv:hep-th/0112044].
  %%CITATION = HEP-TH 0112044;%%

  R.~R.~Metsaev and A.~A.~Tseytlin,
  {\it Exactly solvable model of superstring in plane wave Ramond-Ramond
  background,}
  Phys.\ Rev.\ D {\bf 65} (2002) 126004,
  [arXiv:hep-th/0202109].
  %%CITATION = HEP-TH 0202109;%%

\bibitem{Sprad1} M.~Spradlin and A.~Volovich,
  {\it Superstring interactions in a pp-wave background,}
  Phys.\ Rev.\ D {\bf 66} (2002) 086004,
  [arXiv:hep-th/0204146].
  %%CITATION = HEP-TH 0204146;%%

\bibitem{Huang} M.-x.~Huang,
  {\it Three point functions of N = 4 super Yang Mills from light cone string
  field theory in pp-wave,}
  Phys.\ Lett.\ B {\bf 542} (2002) 255,
  [arXiv:hep-th/0205311].
  %%CITATION = HEP-TH 0205311;%%

\bibitem{Chu} C.~S.~Chu, V.~V.~Khoze and G.~Travaglini,
  {\it Three-point functions in N = 4 Yang-Mills theory and pp-waves,}
  JHEP {\bf 0206} (2002) 011,
  [arXiv:hep-th/0206005].
  %%CITATION = HEP-TH 0206005;%%

\bibitem{Sprad2} M.~Spradlin and A.~Volovich,
  {\it Superstring interactions in a pp-wave background. II,}
  JHEP {\bf 0301} (2003) 036,
  [arXiv:hep-th/0206073].
  %%CITATION = HEP-TH 0206073;%%

\bibitem{Schwarz} J.~H.~Schwarz,
{\it Comments on superstring interactions in a plane-wave background,}
JHEP {\bf 0209} (2002) 058,
[arXiv:hep-th/0208179].
%%CITATION = HEP-TH 0208179;%%

\bibitem{Pank} A.~Pankiewicz and B.~J.~Stefanski,
  {\it pp-wave light-cone superstring field theory,}
  Nucl.\ Phys.\ B {\bf 657} (2003) 79,
  [arXiv:hep-th/0210246].
  %%CITATION = HEP-TH 0210246;%%

\bibitem{He} Y.~H.~He, J.~H.~Schwarz, M.~Spradlin and A.~Volovich,
  {\it Explicit formulas for Neumann coefficients in the plane-wave geometry,}
  Phys.\ Rev.\ D {\bf 67} (2003) 086005,
  [arXiv:hep-th/0211198].
  %%CITATION = HEP-TH 0211198;%%

\bibitem{Lucietti1} J.~Lucietti, S.~Schafer-Nameki and A.~Sinha,
  {\it On the exact open-closed vertex in plane-wave light-cone string field
  theory,}
  Phys.\ Rev.\ D {\bf 69} (2004) 086005,
  [arXiv:hep-th/0311231].
  %%CITATION = HEP-TH 0311231;%%

\bibitem{Lucietti2} J.~Lucietti, S.~Schafer-Nameki and A.~Sinha,
  {\it On the plane-wave cubic vertex,}
  Phys.\ Rev.\ D {\bf 70} (2004) 026005,
  [arXiv:hep-th/0402185].
  %%CITATION = HEP-TH 0402185;%%


\bibitem{Lucietti3} J.~Lucietti,
  {\it A note on the uniqueness of the Neumann matrices in the plane-wave
  background,}
  Phys.\ Lett.\ B {\bf 598} (2004) 285,
  [arXiv:hep-th/0407035].
  %%CITATION = HEP-TH 0407035;%%
 
\bibitem{Maldacena}
  J.~M.~Maldacena,
  {\it TASI 2003 lectures on AdS/CFT,}
  [arXiv:hep-th/0309246].
  
\bibitem{Sadri}
  D.~Sadri and M.~M.~Sheikh-Jabbari,
  {\it The plane-wave / super Yang-Mills duality,}
  Rev.\ Mod.\ Phys.\  {\bf 76} (2004) 853,
  [arXiv:hep-th/0310119].
  %%CITATION = HEP-TH 0310119;%%

\bibitem{Russo}
  R.~Russo and A.~Tanzini,
  {\it The duality between IIB string theory on pp-wave and N = 4 SYM: 
  A  status report,}
  Class.\ Quant.\ Grav.\  {\bf 21} (2004) S1265,
  [arXiv:hep-th/0401155].
 
 
\bibitem{RSZ}
  L.~Rastelli, A.~Sen and B.~Zwiebach,
 {\it Classical solutions in string field theory around the tachyon vacuum,}
  Adv.\ Theor.\ Math.\ Phys.\  {\bf 5} (2002) 393,
  [arXiv:hep-th/0102112].
  %%CITATION = HEP-TH 0102112;%%

\end{thebibliography}
\end{document}